\documentclass[preprint,12pt,5p,authoryear]{elsarticle}
\usepackage[utf8]{inputenc}

\makeatletter
\def\ps@pprintTitle{%
 \let\@oddhead\@empty
 \let\@evenhead\@empty
 \def\@oddfoot{}%
 \let\@evenfoot\@oddfoot}
\makeatother

\usepackage{amssymb}
\usepackage{amsmath}

\usepackage{amsthm}

\usepackage{todonotes}

\usepackage{xcolor}

\newcommand{\blue}[1]{\textcolor{black}{#1}}

\usepackage[colorlinks=true,menucolor=.]{hyperref}

\usepackage{microtype} 

\theoremstyle{definition}
\newtheorem{definition}{Definition}

\journal{Annual Reviews in Control}

\begin{document}

\begin{frontmatter}
\title{An Overview on Optimal Flocking}

\author{Logan E. Beaver, Andreas A. Malikopoulos}
\address{Department of Mechanical Engineering, University of Delaware, Newark, DE, 19716, USA}

\begin{abstract}
The study of robotic flocking has received considerable attention in the past twenty years.
As we begin to deploy flocking control algorithms on physical multi-agent and swarm systems, there is an increasing necessity for rigorous promises on safety and performance.
In this paper, we present an overview the literature focusing on optimization approaches to achieve flocking behavior that provide strong safety guarantees.
We separate the literature into cluster and line flocking, and categorize cluster flocking with respect to the system\blue{-level} objective, which may be realized by a reactive or planning control algorithm.
\blue{We also categorize the line flocking literature by the energy-saving mechanism that is exploited by the agents.}
We present several approaches aimed at minimizing \blue{the} communication and computational requirements in real systems via neighbor filtering and event-driven planning\blue{, and} conclude with our perspective on the outlook and future research direction of optimal flocking \blue{as a field}.
\end{abstract}



\end{frontmatter}

{
\hypersetup{linkcolor=black}
\setcounter{tocdepth}{2}
\tableofcontents
}

\section{Introduction} \label{sec:introduction}

Generating emergent flocking behavior has been of particular interest since Reynolds proposed three heuristic rules for multi-agent flocking in computer animation; see \cite{Reynolds1987}. Robotic flocking has been proposed in several applications including mobile sensing networks, coordinated delivery, reconnaissance, and surveillance; see \cite{Olfati-Saber2006FlockingTheory}.
With the significant advances in computational power in recent decades, the control of robotic swarm systems has attracted considerable attention due to their adaptability, scalability, and robustness to individual failure; see \cite{Oh2017}.
However, constructing a swarm with a large number of robots imposes significant cost constraints on each individual robot.
Thus, any real robot swarm \blue{must} consist of individual robots with limited sensing, communication, actuation, memory, and computational abilities.
To achieve robotic flocking in a physical swarm, we must develop and employ energy-optimal approaches to flocking under these strict cost constraints.

There have been several surveys and tutorials on decentralized control that include flocking; see 
\cite{Barve,Bayindir2016,Ferrari2016DistributedTutorial,Zhu2016RecentSurvey,Zhu2017ASystems,Albert2018Survey:Tracking}.
In one motivating example, \cite{Fine2013UnifyingMembers} discuss various flocking controllers without considering optimality.
In general, these surveys have all considered flocking and optimal control to be two distinct problems.
Thus, we believe it is appropriate to present a comprehensive overview of optimal flocking control algorithms as robotic swarm systems begin to roll out in laboratories, e.g.,  \cite{Rubenstein2012,jang2019simulation,Beaver2020DemonstrationCity,Malikopoulos2020,Wilson2020TheSystems}, and field tests, e.g.,  \cite{Vasarhelyi2018OptimizedEnvironments,mahbub2020sae-1}.
Our objective for this overview is to establish the current frontier of optimal flocking research and present our vision of the research path for the field.

While an extensive body of literature has studied the convergence of flocking behavior, there has been almost no attention \blue{paid} to the development of optimal flocking control algorithms.
Although \cite{Molzahn2017ASystems} focused on optimal decentralized control in a recent survey, the \blue{covered} approaches focus on formation configuration\blue{,} achieving consensus, or area coverage.
\blue{Thus,} we seek to summarize the existing literature at the interface of flocking and optimization with an emphasis on minimizing agents' energy consumption.

\begin{figure}[ht]
    \centering
    \includegraphics[width=0.95\linewidth]{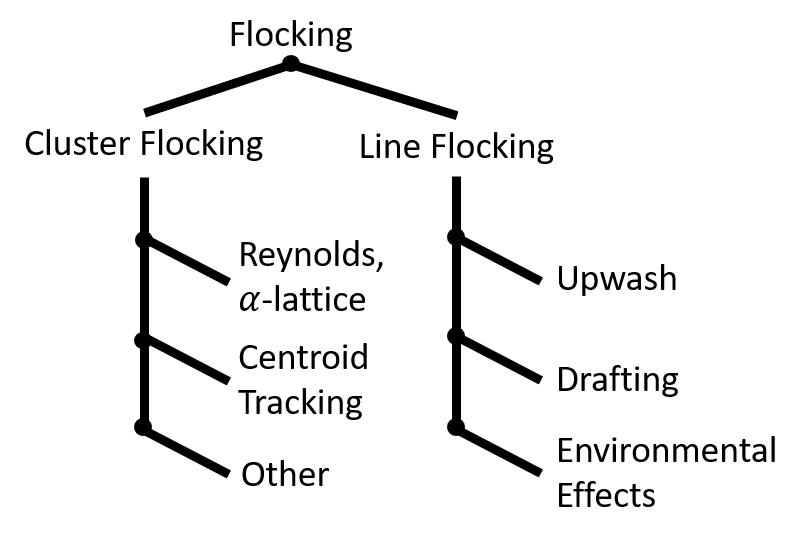}
    \caption{Our proposed flocking classification scheme for cluster and line flocking. }
    \label{fig:flockingCategories}
\end{figure}

One significant problem throughout the literature is the use of the term ``flocking'' to describe very different modes of aggregate motion.
The biology literature emphasizes this point, e.g., \cite{Bajec2009OrganizedBirds}, where the distinction of line flocking (e.g., geese) and cluster flocking (e.g., sparrows) is necessary.
To this end, we believe it is helpful to partition the engineered flocking literature into cluster and line flocking.
As with natural systems, these modes of flocking have vastly different applications and implementations; unlike biological systems, the behavior of engineering systems is limited only by the creativity of the designer.
Our proposed flocking taxonomy is shown in Fig. \ref{fig:flockingCategories}.
\blue{
We have  partitioned cluster flocking into three categories based on the desired system-level objective.
In the context of optimal flocking, each of our proposed categories corresponds to a system-level cost function that a designer seeks to minimize by optimizing the control policy of each agent.
In contrast, the objective of line flocking is always to minimize the energy consumed by each agent.
Thus, we have partitioned the line flocking literature based on the mechanism that agents exploit to reduce energy consumption.
}

The objectives of this paper are to: 
(1) elaborate on a new classification scheme for engineered flocking literature aimed at enhancing the description of flocking research (Fig. \ref{fig:flockingCategories}), 
(2) summarize the results of the existing optimal flocking literature across engineering disciplines and present the frontier of \blue{optimal} flocking research, and 
(3) propose a new paradigm to understand flocking as an emergent phenomenon to be controlled rather than a desirable group behavior for agents to mimic.

The contribution of this paper is the collection and review of the literature in this area. In several cases the optimal flocking and formation 
literature overlap. We have attempted to separate them and present only the material relevant to flocking in this review. Any such effort has obvious limitations. Space constraints limit the presentation of technical details, and thus, extensive discussions are included only where they are important for understanding the fundamental concepts or explaining significant departures from previous work. 

The remainder of this paper is structured as follows. In Section \ref{ss:notation}, we present the common notation used throughout the paper.
Then, in Section \ref{sec:clusterFlocking}, we give a \blue{brief} introduction to cluster flocking\blue{.}
\blue{
We review the optimal cluster flocking literature in Sections \ref{sec:reynoldsFlocking} - \ref{sec:other} following our proposed taxonomy (Fig. \ref{fig:flockingCategories}).
In Section \ref{sec:reynoldsFlocking}, we present various approaches to minimize the deviation from Reynolds flocking rules.
We present approaches to optimally track a reference trajectory with the center of a flock in Section \ref{sec:refTracking}, and in Section \ref{sec:other}, we present other system-level objectives where optimal control policies induce cluster flocking.}
We further divide each of these sections into reactive and planning approaches.
We present the line flocking literature in Section \ref{sec:line}\blue{. In } 
Section \ref{sec:pareto}, \blue{we discuss the inherent trade-offs present in multi-objective optimal flocking through the lens of Pareto analysis.  }
In Section \ref{sec:physical}, we discuss the implications of flocking with real robots.
In Section \ref{sec:cyberPhysical}, we present approaches to reducing cyberphysical costs, while in Section \ref{sec:strategy} we present flocking as a group strategy.
Finally, in Section \ref{sec:outlook}, we present our research outlook, concluding remarks, and motivate a new direction for flocking research.

\subsection{Notation} \label{ss:notation}

We consider a swarm of $N\in\mathbb{N}$ agents indexed by the set $\mathcal{A} = \{1, 2, \dots, N\}$.
For each agent $i\in\mathcal{A}$, we denote their position and velocity by $\mathbf{p}_i(t)$ and $\mathbf{v}_i(t)$, respectively, at time $t\in\mathbb{R}_{\geq0}$.
Agent $i$'s state is denoted by the vector $\mathbf{x}_i(t)$, and the state of the system by $\mathbf{x}(t) = [\mathbf{x}_1^T(t), \dots, \mathbf{x}_N^T(t)]^T$. 
Each agent $i\in\mathcal{A}$ has a neighborhood $\mathcal{N}_i(t) \subseteq \mathcal{A}$, which contains all neighbors that $i$ may sense and communicate with. For consistency we explicitly include $i\in\mathcal{N}_i(t)$ for all $t$.
There are many ways to define a neighborhood, including inter-agent distance, $k$-nearest neighbors, and Voronoi partitions; see \cite{Fine2013UnifyingMembers} for further discussion.
In most cases, each agent's neighborhood is only a fraction of the swarm; thus, each agent is only able to make partial observations of the entire system state.
Using neighborhoods as our basis for local information, we propose the following definition for connected agents.
\begin{definition} \label{def:connected}
Two agents $i,j\in\mathcal{A}$ are connected at time $t$ over a period $T\in\mathbb{R}_{>0}$ if there exist a sequence of neighborhoods
\begin{equation}
    \big\{\mathcal{N}_i(t_1), \mathcal{N}_{k_1}(t_2), \mathcal{N}_{k_2}(t_3), \dots, \mathcal{N}_{k_n}(t_{n+1})\big\},
\end{equation}
such that
\begin{equation}
    k_1 \in \mathcal{N}_i(t_1), k_2 \in \mathcal{N}_{k_1}(t_2), \dots, j\in\mathcal{N}_{k_{n}}(t_{n+1}),
\end{equation}
where $n+1$ is the length of the sequence and $t\leq t_1 \leq t_2 \dots \leq t_{n+1} \leq t + T$.
\end{definition}

Finally, for any two agents $i, j \in\mathcal{A}$, we denote their relative position as
\begin{equation}
    \mathbf{s}_{ij}(t) = \mathbf{p}_{\blue{i}}(t) - \mathbf{p}_{\blue{j}}(t).
\end{equation}

\section{Cluster Flocking and Swarming} \label{sec:clusterFlocking}

The swarming, aggregate motion of small birds is known as cluster flocking in biological literature.
The benefit of cluster flocking in natural systems is unknown, however, several hypotheses have been proposed.
These include predator evasion, estimating the flock population, and sensor fusion.
It is also unclear if leadership is necessary to generate the organized motion in cluster flocks of actual birds;  \cite{Bajec2009OrganizedBirds} provides a review of swarming in biological systems.
\blue{
In fact, the conditions under which leader-driven cluster flocking is optimal is an open question; \cite{Jia2019ModellingFlocking} provides further results on hierarchical cluster flocking.
For this reason, the study of leader vs leaderless flocking is outside the scope of this review.
}

In \blue{Sections \ref{sec:reynoldsFlocking} - \ref{sec:other}}, we present each formulation \blue{under the assumption that} all agents have access to any global reference information when solving their local optimization problem.
With this in mind, and based on the work of \cite{Olfati-Saber2006FlockingTheory,Cucker2007EmergentFlocks,Tanner2007}, we present a general definition for cluster-flocking behavior in engineered swarms.

\begin{definition}{(Cluster Flocking)} \label{def:clusterFlocking}
A group of agents achieves cluster flocking if:
\begin{enumerate}
    \item There exists a finite distance $D\in\mathbb{R}_{>0}$ such that $||\mathbf{p}_i(t) - \mathbf{p}_j(t)|| \leq D$ for all $i, j\in\mathcal{A}$ and all $t\in\mathbb{R}_{\geq0}$.
    \item There exists a finite period of time $T\in\mathbb{R}_{>0}$ such that every pair of agents $i,j \in\mathcal{A}$ is connected for all $t\in\mathbb{R}_{\geq0}$ (Definition \ref{def:connected}).
    \item No agent $i\in\mathcal{A}$ has a desired final state (i.e., there is no explicit formation).
    \item \blue{The agents do not remain stationary}.
\end{enumerate}
\end{definition}

The first component of Definition \ref{def:clusterFlocking} draws from the idea of cohesion in \cite{Olfati-Saber2006FlockingTheory} and \cite{Cucker2007EmergentFlocks}, where the flock must stay within some finite bounded diameter. The second component of Definition \ref{def:clusterFlocking} is inspired by \cite{Jadbabaie2003}, which shows that agents can converge to velocity consensus even when the communication topology of the flock is only connected over time.
The third component of Definition \ref{def:clusterFlocking} seeks to differentiate cluster flocking from formation control problems, which imposes an explicit structure on the agents.
Finally, the fourth component of Definition \ref{def:clusterFlocking} requires the flock to remain mobile, which is necessary to differentiate flocking from area coverage.

\blue{
In almost all cluster flocking applications, each individual agent $i\in\mathcal{A}$ may only observe neighboring agents $j\in\mathcal{N}_i(t)$ at any particular instant in time.
In general $|\mathcal{N}_i(t)| < |\mathcal{A}|$, and therefore each agent may only make partial observations of the system state, $\mathbf{x}(t)$.
Thus, many of the approaches in the optimal cluster flocking literature rely on repeatedly simulating the system, evaluating the global cost, and updating the agents' control policies.
}

\blue{
Sections \ref{sec:reynoldsFlocking}, \ref{sec:refTracking}, and \ref{sec:other} each correspond to distinct system-level objectives where cluster flocking emerges.
In each section we present literature and key results for optimizing the control policy of each agent such that the system-level cost is minimized.
We further split each section into subsections on reactive and planning optimization approaches.
}

\section{Reynolds Flocking} \label{sec:reynoldsFlocking}

\begin{figure}[ht]
    \centering
    \includegraphics[width=.7\linewidth]{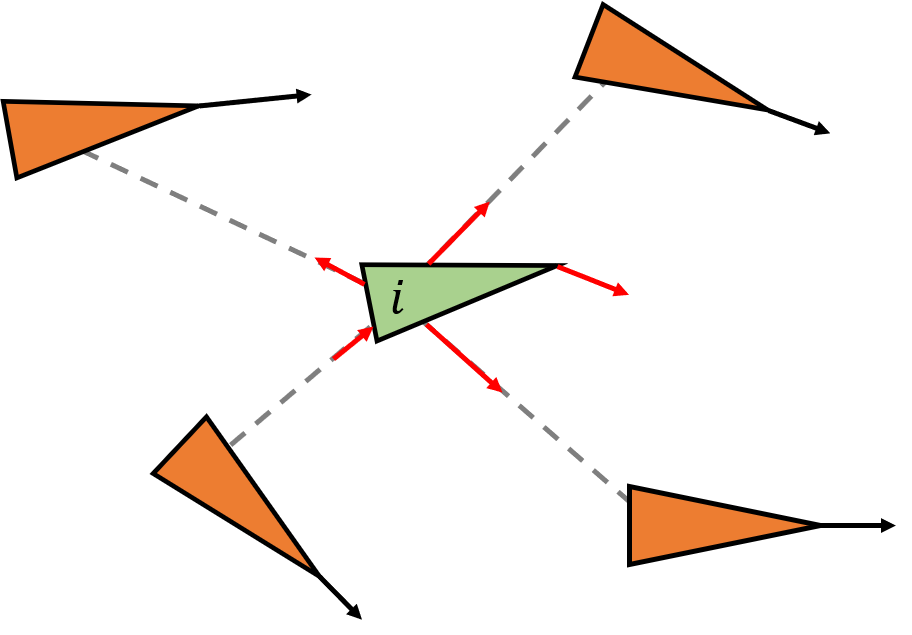}
    \caption{A diagram showing the influence of collision avoidance, flock centering, and velocity matching for agent $i$, in green.}
    \label{fig:reynolds}
\end{figure}

A vast amount of literature exists that seeks to achieve flocking under Reynolds \blue{three flocking rules, (1) collision avoidance, (2) velocity matching, and (3) flock centering; see \cite{Reynolds1987}}.
Generally \blue{these heuristic rules} can be captured by imposing \blue cost function \blue{of the following form} for each agent $i\in\mathcal{A}$,
\begin{equation} \label{eq:reynoldsCost}
    J_i = V(||\mathbf{s}_{ij}(t)||) + \sum_{j\in\mathcal{N}_i(t)}||\dot{\mathbf{s}}_i(t)||^2,
\end{equation}
where $j\in\mathcal{N}_i(t)$ and $V$ is an attractive-repulsive potential function with a local minimum at some desired distance.
The first term of \eqref{eq:reynoldsCost} manages collision avoidance and flock centering, while the second term ensures velocity alignment. Fig. \ref{fig:reynolds} shows each component of an agent flocking under Reynolds rules.

Given a distance $d\in\mathbb{R}_{>0}$ that minimizes the potential function in \eqref{eq:reynoldsCost}, \cite{Olfati-Saber2006FlockingTheory} proposes the $\alpha$-lattice, i.e.,  any configuration of agents such that each agent $i\in\mathcal{A}$ satisfies
\begin{equation} \label{eq:alphaLattice}
    ||\mathbf{s}_{ij}(t)|| = d,
\end{equation}
for all $j\in\mathcal{N}_i(t)$. This definition coincides with the global minimum of \eqref{eq:reynoldsCost}, and many authors substitute \eqref{eq:alphaLattice} for the flock centering and collision avoidance rules of Reynolds.
\blue{In the following subsections, we present reactive and planning approaches to minimize each agent's deviation from Reynolds flocking rules.}

\subsection{Reactive Approaches}


An early approach to optimally follow Reynolds flocking rules \blue{is} presented by \cite{Morihiro2006EmergenceLearning}, where the authors \blue{take} a learning-based approach to velocity alignment.
In this work, each agent $i\in\mathcal{A}$ observes the state, $\mathbf{x}_j(t)$, of a randomly selected agent $j\in\mathcal{A}\setminus\{i\}$ at each time step $t$.
Agent $i$ then follows one of four motion primitives: (1) move toward $j$, (2) move away from $j$, (3) move the same direction as $j$, or (4) move the opposite direction of $j$.
The agents are rewarded for achieving velocity alignment and staying near some desirable distance $d$ of their neighbors, i.e., velocity matching and flock centering.
In addition, the authors include a set of predators that attempt to disrupt the flock.
In this case, the agents observe the state of the predator with probability $1$ whenever it is within range.
Agent $i$ is rewarded for evading the predator and maintaining the structure of the flock.
Further simulation results for this method are presented in \cite{Morihiro2006CharacteristicsScheme}.

\cite{Wang2018AEnvironments} \blue{formulates flocking as a dynamic program} to generate optimal trajectories for a swarm of quadrotors in $\mathbb{R}^2$.
The \blue{system-level} objective is for the quadrotors to follow Reynolds flocking rules while moving the swarm center to a global reference position.
The agents follow unicycle dynamics, and each agent\blue{'s neighborhood consists of its} nearest left and right neighbor. 
This angular symmetry 
reduces the likelihood of \blue{any small subset of agents } forming \blue{an} isolated clique, which \blue{can easily occur when agents use distance-based and nearest-neighbor rules to define their neighborhood}; 
see \cite{Camperi2012SpatiallyModels,Fine2013UnifyingMembers}.
The authors penalize each agent for violating Reynolds flocking rules, coming within some distance of an obstacle, \blue{or} not moving toward the desired location.
They also incorporate a constant transition penalty if the agent is not within a fixed distance of the goal, \blue{which encourages the agent to quickly approach it}.
\blue{The authors follow a standard deep reinforcement learning policy, which they }
verify on a group of $N=3$ agents \blue{in simulation}.
\blue{They also} show that the decentralized control policy generalizes to larger systems of $5$ and $7$ uncrewed aerial vehicles (UAVs) without significant deterioration of the final objective function value.

Metaheuristic algorithms, including Pigeon-inspired optimization, \blue{proposed by} \cite{Duan2014Pigeon-inspiredPlanning}, and particle swarm optimization, \blue{proposed by} \cite{Kennedy1995ParticleOptimization}, have been used to generate systems that optimally follow Reynolds flocking rules.
In \cite{Qiu2020AObstacles}, the authors \blue{optimally select} the control actions of a UAV in $\mathbb{R}^3$ \blue{under} state and control constraints.
This is achieved by breaking the controller into flocking and obstacle avoidance components, then using pigeon-inspired optimization to weigh each component such that the deviation from Reynolds flocking rules \blue{is} minimized while avoiding collisions.

\cite{Navarro2015DistributedBehaviors} applie\blue{s} particle swarm optimization to a neural network controller with $50$ weights, nine inputs, and two outputs.
The inputs consist of distance measurements for each octant around the agent and the average heading of all neighboring agents.
The outputs of the neural network are speed commands for the left and right motor of a differential drive robot.
The system is trained to maximize a linear combination of local velocity alignment, desired inter-robot spacing, and the average velocity of the flock.
The agents are trained in simulation in the local and global information case.
The authors \blue{demonstrate} that a neural network trained on $4$ agents can generalize up to groups of \blue{at least} $16$.

The effect of control input constraints for an optimal flocking controller \blue{is} studied in \cite{Celikkanat2008OptimizationStrategies}.
In this work, the authors \blue{seek} to design a local control law \blue{to simultaneously} maximiz\blue{e} velocity alignment \blue{and minimize} 
deviation from an $\alpha$-lattice.
\blue{The authors propose two global parameters that the agents may access: the average heading of all agents, and an \emph{order parameter} based on }
Shannon's information entropy metric, \blue{first proposed in  \cite{Shannon1948ACommunication}}, \blue{which considers} the proportion of robots within a disk of diameter $h$.
\blue{The authors employ a genetic algorithm to find the optimal control parameters, and the algorithm's }
performance is validated in simulation.
Another genetic algorithm \blue{is} employed by \cite{Vasarhelyi2018OptimizedEnvironments} to design \blue{a shared} feedback controller \blue{employed by each} individual agent, which is parameterized in terms of $11$ optimization variables.
The authors optimize the agents in a constrained environment with a complicated objective function that includes the minimization of collision risk with walls and other agents, deviation from desired speed, and the number of disconnected agents, while simultaneously maximizing velocity alignment and the largest cluster size.
The control variables are optimized offline in a realistic simulation that includes stochastic disturbances for desired flock speeds of $4$, $6$, and $8$ m/s.
The controller is validated in outdoor flight experiments with 30 Pixiehawk drones flown over 10-minute intervals.

\blue{Up to this point, obstacle avoidance and safety have primarily been achieved through the use of }
artificial potential fields and attractive-repulsive forces\blue{, e.g., by minimizing \eqref{eq:reynoldsCost}}.
In addition, the design of potential fields has been the subject of significant research for general navigation problems; see \cite{Vadakkepat2000EvolutionaryPlanning}.
However, \blue{\cite{Koren1991PotentialNavigation} have demonstrated several major drawbacks associated with potential field methods. }
These include introducing steady oscillations to trajectories and exacerbating deadlock in crowded environments.

A promising alternative to potential field methods, which explicitly guarantees safety, has been proposed as a novel paradigm for the design of long-duration robotic systems by \cite{Egerstedt2018RobotAutonomy}. In this approach, the tasks of each agent are imposed as motion constraints while the agents seek energy-minimizing trajectories.
We interpret this constraint-driven approach to control as understanding why agents take particular control actions, rather than designing control algorithms that mimic a desirable behavior.
To the best of our knowledge, reactive constraint-driven Reynolds flocking has only been explored by \cite{Ibuki2020Optimization-BasedBodies}. Under this approach, each agent $i\in\mathcal{A}$ generates an optimal control trajectory by solving the following optimal control problem at each time $t$,
\begin{align}
    &\min_{\mathbf{u}_i(t)\in\mathbb{R}^6, \delta_i\in\mathbb{R}} ||\mathbf{u}_i(t)||^2 + \delta_i^2 \notag\\
    \text{subject to:} \notag\\
    &\lim_{t\to\infty} ||\mathbf{s}_{ij}(t)|| \leq \delta_i, \label{eq:pose}\\
    &\lim_{t\to\infty} ||\phi_{ij}(t)|| \to 0, \label{eq:velSync}\\
    &||\mathbf{s}_{ij}(t)|| > 2R \quad \forall t\in\mathbb{R}_{\geq0}, \label{eq:colAv}\\
    &\quad\forall j\in\mathcal{A}\setminus\{i\}, \notag
\end{align}
where $\delta$ is a slack variable,  $\phi_{ij}$ is a metric for attitude error, and $R$ is the radius of a \blue{safety} circle that circumscribes the agents. Constraint \eqref{eq:pose} corresponds to pose synchronization (flock centering), \eqref{eq:velSync} to attitude synchronization (velocity alignment), and \eqref{eq:colAv} to collision avoidance.
The authors generate control inputs for each agent by applying gradient flow coupled with \blue{a} control barrier function to \blue{guarantee} constraint satisfaction.
\blue{Thus, under the constraint-driven approach,} the agents satisfy the safety constraint \blue{and realize} Reynolds flocking rules within a threshold $\delta$ \blue{ while traveling along a trajectory that minimizes the energy consumption of each individual agent.}

\subsection{Planning Approaches}

As an alternative to simply reacting to the \blue{current state of the} environment and system, agents may instead plan an optimal trajectory over a time horizon.
This can generally improve the performance of the agent, e.g., by avoiding local minima; however, planning generally requires more computational power than a reactive approach.
The structure of the information in a decentralized system also creates challenges with respect to the information available over a planning horizon.
It has been shown that there is separation between estimation and control for particular decentralized information structures; see \cite{nayyar2013decentralized,Dave2020a}. However, this is an open problem for the general case.
Some proposed solutions include sharing information between agents, e.g., see \cite{Morgan2016}, only planning with agents shared between neighbors, e.g., see \cite{Dave2019a}, and applying model predictive control (MPC).
For large swarms of inexpensive agents, widespread information sharing is generally infeasible, and it is unlikely that any common information exists.
For this reason, MPC has been a preferred approach in swarm systems.
With MPC, each agent plans a sequence of control actions over a time horizon based on its current information about the system.
After some time, the agent will replan its trajectory based on whatever new information it has received.

A significant number of MPC flocking algorithms seek to minimize deviation from Reynolds flocking rules, which may be implemented through a linear combination of the following objectives:
\begin{align}
        J_i^d(t) &= \sum_{j\in N_i(t)} \Big(||\mathbf{s}_{ij}(t)|| - d\Big)^2, \label{eq:distanceCost} \\
        J_i^v(t) &= ||\bar{\mathbf{v}}_i(t) - \mathbf{v}_i(t)||^2, \label{eq:velocityCost} \\
        J_i^u(t) &=||\mathbf{u}_i(t)||^2, \label{eq:controlCost}
\end{align}
where $d$ is the desired separating distance in \eqref{eq:alphaLattice}, and $\bar{\mathbf{v}}_i(t)$ is the average velocity of all agents $j\in\mathcal{N}_i(t)$. Eq. \eqref{eq:distanceCost} corresponds to flock centering, \eqref{eq:velocityCost} to velocity matching, and \eqref{eq:controlCost} is a control effort penalty term.

The analysis by \cite{Zhang2008CollectiveMechanisms} presents a mechanism for flocking agents to estimate their neighbors' future trajectories. The predictive device \blue{is} applied by \cite{Zhan2011FlockingMechanisms} to achieve Reynolds flocking under a fully connected communication topology. This \blue{is} extended to the decentralized information case in \cite{Zhan2011DecentralizedMechanisms} and validated experimentally with outdoor flight tests in \cite{Yuan2017OutdoorControl}.

An infinite horizon continuous-time MPC approach \blue{is} employed in \cite{Xu2015DistributedSystems} and \cite{Xu2017FastSystem} \blue{to} minimize flocking error over an infinite horizon in a continuous-time system.
The resulting Hamilton-Jacobi-Bellman equation is nonlinear and \blue{does not have} a \blue{known} explicit solution.
As a result, the authors apply an original reinforcement learning technique to optimize the agent trajectories online and validate the performance in simulation.
The reinforcement learning architecture is expanded on in \cite{Jafari2020AEnvironment}, where the authors include model mismatch and significant environmental disturbances acting upon the agents.
They also present simulation and experimental results for a flock of quadrotors \blue{moving through $\mathbb{R}^3$}.

To guarantee feasibility of the planned trajectories, it is necessary to explicitly impose constraints that bound the maximum control and velocity of each individual agent within their physical limits, e.g., for each agent $i\in\mathcal{A}$,
\begin{align}
    ||\mathbf{v}_i(t)|| &\leq v_{\max}, \label{eq:vConstraint} \\
    ||\mathbf{u}_i(t)|| &\leq u_{\max}, \label{eq:uConstraint}
\end{align}
for all $t\in\mathbb{R}_{\geq0}$.
An analysis of constrained $\alpha$-lattice flocking under MPC, which incorporates \eqref{eq:vConstraint} and \eqref{eq:uConstraint}, \blue{is} explored in \cite{Zhang2015ModelConstraints}, and \blue{is} extended to velocity alignment in \cite{Zhang2016ModelConstraints}.

As we begin to implement flocking in physical swarms, explicit guarantees of safety are imperative for any proposed control algorithm. The most straightforward approach to guarantee safety is to circumscribe each agent $i\in\mathcal{A}$ entirely within a closed disk of radius $R\in\mathbb{R}_{>0}$. The safety constraint for $i$ may then be formulated as
\begin{equation} \label{eq:safety}
    ||\mathbf{s}_{ij}(t)|| \geq 2R, \quad \forall j\in\mathcal{A}\setminus\{i\}.
\end{equation}

In general, applying MPC to each agent does not guarantee that coupled constraints, such as \eqref{eq:safety}, are satisfied.
At any planning instant, agent $i$ only has the trajectories generated by $j\in\mathcal{N}_i(t)$ at a previous time step.
Thus, agent $i$ \blue{is unable to determine whether \eqref{eq:safety} is satisfied for the unknown current trajectory of $j$}. 
For this reason, in the decentralized case, agents must either cooperatively plan trajectories or impose a compatibility constraint.
To guarantee that coupled constraints between agents are satisfied, significant research effort has been dedicated to decentralized MPC (DMPC). A common approach to DMPC is to design a communication protocol for agents to iteratively generate trajectories while driving their cost to a local minimum.
An iterative approach, proposed by \cite{Zhan2013FlockingMeasurements}, cooperatively generates trajectories while limiting the number of messages exchanged between agents. The agents apply an impulse acceleration at discrete intervals and seek to minimize the flock centering error over a finite horizon. The agents sequentially generate trajectories up to some index $l\geq N$, where at each iteration, agent $i = \big(k \mod N\big) + 1, ~ i\in\mathcal{A}, ~ k = 0, 1, \dots, l-1$ generates its trajectory.
This guarantees that the coupled safety constraints are satisfied and that the cost of agent $i$'s trajectory is nonincreasing with each planning iteration.

\blue{In contrast,} \cite{Beaver2020AnFlocking} \blue{takes a cooperative approach to DMPC, where } Reynolds flocking rules \blue{are taken} as an endpoint cost in a continuous optimal control problem while including \eqref{eq:vConstraint} - \eqref{eq:safety} as constraints. Each agent $i\in\mathcal{A}$ first generates a trajectory while relaxing the safety constraint \eqref{eq:safety}. Agent $i$ then exchanges trajectory information with every other $j\in\mathcal{N}_i(t)$. Finally, any agents violating \eqref{eq:safety} cooperatively generate the centralized safety-constrained trajectory between fixed start and end points, which guarantees safety\blue{, state, and control constraint satisfaction}.

\section{Reference State Cluster Flocking} \label{sec:refTracking}

\begin{figure}[ht]
    \centering
    \includegraphics[width=0.7\linewidth]{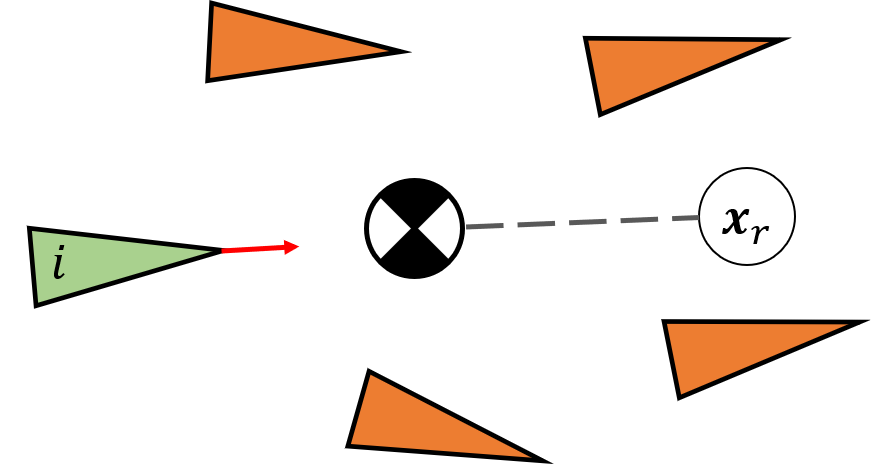}
    \caption{Agent $i$, in green, selects the control input that drives the center of the flock toward the reference state, $\mathbf{x}_r$.}
    \label{fig:referenceCluster}
\end{figure}

A common application \blue{where cluster flocking emerges} 
is tracking a reference trajectory with the center of mass of a swarm.
In this application, the reference trajectory (also called a virtual leader) is generally presented as a time-varying reference state, $\mathbf{x}_r(t)$, which may be known to all agents.
In general, this is \blue{by including an additional cost to \eqref{eq:reynoldsCost} with the form} 
\begin{equation} \label{eq:comFeedback}
    J_i^r(t) = \Bigg|\Bigg| \blue{ \frac{1}{N} \sum_{i\in\mathcal{A}} }\mathbf{x}_i(t) - \mathbf{x}_r(t)  \Bigg|\Bigg|^2,
\end{equation}
which may be \blue{weighted with a positive scalar}. 
\blue{The system-level objective of the swarm is to track the reference state, $\mathbf{x}_r(t)$, within some threshold $\epsilon\in\mathbb{R}_{\geq0}$, i.e.,}
\begin{equation} \label{eq:comObjective}
\Bigg|\Bigg|\frac{1}{N} \sum_{i\in\mathcal{A}} \big(\mathbf{x}_i(t)\big) - \mathbf{x}_r(t) \Bigg|\Bigg| \leq \epsilon.
\end{equation}
As with Reynolds flocking, the information available to each agent $i\in\mathcal{A}$ is restricted to its neighborhood, $\mathcal{N}_i(t)$.
This is\blue{, in general,} insufficient \blue{for any agent to} to evaluate \eqref{eq:comObjective}.
Thus, the center of mass tracking problem has generally been formulated as an optimal controller design problem\blue{, or alternatively each agent might only consider agents within its own neighborhood when evaluating \eqref{eq:comFeedback}}.
A schematic of reference state cluster flocking agents is presented in Fig. \ref{fig:referenceCluster}\blue{, and we present reactive and planning approaches to solve this problem in the following sections}.

\subsection{Reactive Approaches}

An early approach proposed by \cite{Hayes2002Self-OrganizedRobots} \blue{seeks} to track a reference point with a flock of agents that follow Reynolds flocking rules with an additional attractive force \blue{pointing} toward the reference state.
The agents are placed in a rectangular domain, where each agent has a uniform probability of failing over a given period, i.e., the agent would stop moving but still be detectable.
\blue{The authors select the control parameters} that minimize a combination of travel time, cumulative distance traveled, and average inter-agent spacing, \blue{using a standard reinforcement technique, in simulation}.
The resulting controller is validated in physical experiments with $10$ robots.
This \blue{system-level} objective has become standard in many flocking applications; see \citep{Bayindir2016} \blue{for examples}.

Another approach to reference tracking, proposed by \cite{La2009OptimalTracking}, involves \blue{selecting} optimal controller \blue{parameters} such that the reference trajectory $\mathbf{x}_r(t)$ is tracked in minimum time while maintaining an $\alpha$-lattice configuration. The authors construct a cost function that penalizes the time taken for the flock to catch the reference trajectory scaled by their initial position.
The resulting cost function is non-convex and non-differentiable, thus \blue{the authors employ a} genetic algorithm \blue{to minimize it}.
To guarantee that \eqref{eq:comObjective} is globally satisfied, all \blue{controllers} that do not yield an $\alpha$-lattice within some error bound are discarded.
The discrete-time version of this system is optimized by \cite{Khodayari2016FlockingAlgorithm} using a gravitational search algorithm.

\cite{La2015MultirobotAvoidance} propose\blue{s} a \blue{multi-level} flocking \blue{and} learning system to guarantee flocking behavior in the presence of obstacles and predators.
At the \blue{lower} level, each agent seeks to reach the static reference position $\mathbf{p}_r$ with the center of their local neighborhoods.
\blue{At the upper level, each agent must select}, 
in a decentralized way\blue{,} 
the optimal $\mathbf{p}_r \in \mathcal{P}$ from a finite set of positions, $\mathcal{P}$.
Each agent is rewarded proportionately to the size of its neighborhood at each time step, up to a maximum value of 6.
The authors implement a cooperative $Q$-learning approach, where each agent $i\in\mathcal{A}$ \blue{is} rewarded for selecting the appropriate $\mathbf{p}_r$ by
\begin{equation*}
    Q_i^{k+1} = w\, Q_i^k(s_i, a_i) + (1 - w) \sum_{j\in\mathcal{N}_i(t)}Q_j^k(s_j,a_j),
\end{equation*}
where $w\in[0, 1]$ weighs the influence of $i$'s neighbors, and $s_i$, $a_i$ are the state and action taken by agent $i$, respectively.
The convergence properties of this cooperative learning scheme are proved and the performance is demonstrated in simulations and experiments.

To track the reference trajectory under realistic conditions, \cite{Viragh2016Self-organizedEnvironments} \blue{seeks} optimal values for a potential-field based controller in $\mathbb{R}^2$ and $\mathbb{R}^3$ under the effects of sensor noise, communication delay, limited sensor update rate, and constraints on the agent's maximum velocity and acceleration. The work is framed in terms of aerial traffic; thus, multiple competing flocks are placed into shared airspace such that their reference trajectories result in conflict between the flocks.
The authors presented two controllers, one that maintains constant speed and one with a fixed heading.
\blue{In both cases the authors propose a potential field composed of sigmoid functions, which are parameterized by the optimization variables}.
\blue{They construct a} compound objective function \blue{that is} proportional to effective velocity and inversely proportional to collision risk, \blue{and they describe} $20$ scenarios \blue{used} to find $20$ \blue{optimal} parameter sets \blue{for each controller}.
The scenarios consist of every combination of five different initial configurations for both the constant-speed and constant-heading controllers in $\mathbb{R}^2$ and $\mathbb{R}^3$.

As an alternative to deriving an optimal feedback gain, \cite{Atrianfar2013FlockingAgents} \blue{seeks} to minimize the number of informed agents \blue{while guaranteeing that the} entire flock track\blue{s} a known reference trajectory.
First, the authors impose that, for a given sensing distance $h$, the potential field must \blue{approach} infinity as $\mathbf{s}_{ij}$ approaches $h$.
This property guarantees that any connected group of agents remain\blue{s} connected for all time.
Thus, any initially connected group of agents containing an informed agent is guaranteed to converge to the reference trajectory.
The latter implies that at most one informed agent \blue{is} required for each group of connected agents.
In addition, as a function of their initial conditions, some uninformed groups may merge with an informed cluster.
Following this reasoning, the authors \blue{propose that the minimum number of informed agents is not more than the number of initial agent clusters.}

Departing from the aforementioned approaches, a centralized approach to tracking a virtual velocity reference \blue{is} rigorously studied in \cite{Piccoli2016OptimalModel} for double-integrator agents in $\mathbb{R}^k$. The authors present a consensus-driven control law \blue{for each agent $i\in\mathcal{A}$, with the form of} Cucker-Smale flocking,
\begin{align}
    \mathbf{u}_i(t) &= \alpha_i\big(\mathbf{v}_r(t) - \mathbf{v}_i(t)\big) +  (1-\alpha_i)\notag\\
    &\cdot\frac{1}{N-1}\sum_{j\in\mathcal{N}_i(t)\setminus\{i\}}||\mathbf{s}_{ij}(t)||\dot{\mathbf{s}}_{ij}(t),
\end{align}
where $\alpha_i\in[0,1]$ weighs the trade-off between consensus and velocity tracking.
The authors \blue{seek} values of $\alpha_i$ such that $\sum_{i\in\mathcal{A}} \alpha_i \leq M, M\in\mathbb{R}_{>0},$ while minimizing the error function
\begin{equation}
    e(t) = \frac{1}{N}\sum_{i=1}^N||\mathbf{v}_i(t) - \mathbf{v}_r(t)||^2,
\end{equation}
over a time interval $[t^0, t^f] \subset \mathbb{R}_{\geq0}$
The optimal values of $\alpha_i$ \blue{are derived} for three cases: (1) instantaneously minimizing $\frac{d e}{dt}$, (2) minimizing the terminal cost $e(t^f)$, and (3) minimizing the integral cost, $\int_{t^0}^{t^f} e(t)\, dt$.
The resulting optimal control analysis implies that, in general, the optimal strategy is to apply the maximum feedback to a few agents before applying moderate feedback to all agents.
This aims \blue{to drive} agents with high variance toward the reference velocity, enhancing the rate of consensus.
The authors also note the presence of \blue{a} dwell time in the terminal cost case, i.e., the optimal strategy includes applying no control input over a nonzero interval of time starting at $t^0$.

\subsection{Planning Approaches}
Optimal planning has several advantages over reactive methods, although it suffers from a handful of challenges related to information structure \blue{as discussed in Section \ref{sec:reynoldsFlocking}}.
As with the reactive methods, the desired reference trajectory is a time-varying function denoted by $\mathbf{x}_r(t)$.
To guarantee that the reference trajectory can be \blue{tracked}, the agents must be capable of evaluating $\mathbf{x}_r(t)$ over their entire planning horizon.
In addition, each agent $i\in\mathcal{A}$ generally must plan under the assumption that their neighborhood, $\mathcal{N}_i(t)$, is invariant.
Relaxing this assumption may require \blue{levels of} information sharing that \blue{are} infeasible for large swarm systems.

\cite{Lee2013ParticleRobots} appl\blue{y} collective particle swarm optimization to generate \blue{the optimal sequence of control actions} for a general cost function.
In their approach, each agent $i\in\mathcal{A}$ performs a particle swarm optimization with $M\in\mathbb{N}$ particles that correspond to possible control inputs of agent $i$. The agents then transmit their $g < M$ best performing particles to all $j\in\mathcal{N}_i(t)$ and iteratively solve their local particle swarm optimization until the planned trajectories converge system-wide.

\cite{Lyu2019MultivehicleControl} \blue{propose to track} a known reference trajectory by generating the \blue{following} virtual state for each agent $i\in\mathcal{A}$,
\begin{equation} \label{eq:averageNStates}
    \mathbf{z}_i(t) = \frac{1}{|\mathcal{N}_i(t)|} \sum_{j\in\mathcal{N}_i} \mathbf{x}_i(t),
\end{equation}
\blue{i.e., $\mathbf{z}_i(t)$ is} the average state of agent $i$'s neighborhood.
Agent $i$ then imposes the constraint
\begin{equation} \label{eq:avgStateConstraint}
    \mathbf{z}_j(t) = \mathbf{x}_r(t), \quad \forall j\in\mathcal{N}_i(t),
\end{equation}
\blue{and generates its optimal sequence of control actions using the alternating direction method of multipliers (ADMM) technique; see \cite{Summers2012DistributedMultipliers} for further details on ADMM.}
Since $i\in\mathcal{N}_j(t)$ \blue{implies} $j\in\mathcal{N}_i(t)$, the components of \eqref{eq:avgStateConstraint} are shared between neighboring agents \blue{and thus are expected to converge}.

Reference tracking under uncertainty \blue{is} explored by \cite{Quintero2013FlockingApproach} to track the position of a mobile ground vehicle with a known trajectory, $\mathbf{x}_r(t)$.
\blue{Under this approach,} the flocking UAVs travel at a constant speed and altitude with stochasticity in their dynamics.
The objective of each agent is to remain within a predefined annulus centered on the ground vehicle.
The cost for agent $i\in\mathcal{A}$ is defined as the signed distance of agent $i$ from the edge of the annulus plus a heading alignment term.
The authors \blue{determine the optimal control actions of each agent through} dynamic programming. 
\cite{Hung2017AEnvironment} \blue{extend this approach} to include external disturbances, and the optimal policy is derived in real time under a reinforcement learning framework.

\section{Other Cluster Flocking} \label{sec:other}

In addition to Reynolds flocking and centroid tracking, several other applications have been shown to induce cluster flocking behavior. Although not widely addressed in the literature, these results demonstrate the breadth of applications \blue{that would benefit from further analysis of cluster flocking.} 
\blue{One motivating example is presented i}n \cite{Vatankhah2009OnlineMethod}\blue{, where} each agent uses local measurements to determine the control input that maximize\blue{s} the velocity of the swarm center via particle swarm optimization.

\blue{As another example, a}nisotropy in the angle between neighboring flockmates \blue{is} proposed as a metric for measuring the quality of a flock of birds by \cite{Ballerini2008InteractionStudy}.
\cite{Makiguchi2010AnisotropyComputation} construct a measure for anisotropy using a projection matrix
\begin{equation} \label{eq:anisoProjection}
    M^{(n)}_{pq} = \frac{1}{N} \sum_{i\in\mathcal{A}} (\hat{\mathbf{s}}_{ik})\cdot\mathbf{p} ~~ (\hat{\mathbf{s}}_{ik})\cdot\mathbf{q},
\end{equation}
where $k$ indexes the $n$'th nearest neighbor of $i,$ and $\mathbf{p}, \mathbf{q} \in \{\hat{x}, \hat{y}, \hat{z} \}$ are vector components of an orthonormal basis for $\mathbb{R}^3$.
Eq. \eqref{eq:anisoProjection} \blue{is} used to calculate normalized anisotropy, denoted by $\gamma\in[0, 1]$, by comparing the eigenvectors of $M_{pq}^{(n)}$ to the average agent velocity.
The author's objective \blue{is} to select the optimal weights for each of Reynolds flocking rules (cohesion, alignment, and separation) to maximize flock anisotropy for the case that $n = 1$ in \eqref{eq:anisoProjection}.
The authors discard any parameter \blue{sets} that \blue{result in} collisions or flock fragmentation and achieve a final anisotropy of $\gamma \approx 0.8$, which is significantly higher than the critical value for flocking to occur, $\gamma = \frac{1}{3}$.

\cite{Veitch2019ErgodicFlocking} employ\blue{s} ergodic trajectories to achieve flocking.
\blue{A trajectory is ergodic if} 
the average position of the agents over time is equal to some spatially distributed probability mass (or density) function.
\blue{The authors decompose a desired probability density function into a finite Fourier series to provide a metric for the ergodicity of a trajectory.}
The proposed control policy for each robot maximizes th\blue{e ergodicity} metric along an agent's trajectory.
\blue{In addition, e}ach agent $i\in\mathcal{A}$ periodically shares its Fourier coefficients with all $j\in\mathcal{N}_i(t)$.
This allows the agents to predict where their neighbors have previously explored while also guaranteeing collision avoidance by the nature of ergodicity.
Finally, to achieve flocking, the authors generate a uniform probability distribution in a closed disk centered on a reference state in $\mathbb{R}^2$.
By construction, this guarantees that all agents will enter the closed disk and remain within it in finite time. \blue{Finally}, by smoothly moving the disk around $\mathbb{R}^2$, the average velocity and centroid of the flock can be precisely controlled.

\blue{
Optimal shepherding, or influencing, of a flock is explored in \cite{Genter2017FlyFlock}.
In this work the author seeks to inject and optimally control influencing agents within an existing flock of agents that obey Reynolds flocking rules.
The system-level objective is to, in an optimal way, steer real flocks of birds to avoid man-made obstacles, particularly airports.
This is achieved by deriving an optimal \emph{plan}, i.e., sequence of control inputs, such that the entire flock becomes aligned to a target angle in a minimum number of time steps; see \cite{Genter2013AdFlock}.
The author also considers different methods of introducing the influencing agents to an existing flock and the performance implications of these approaches; see \cite{,Genter2015DeterminingFlock,Genter2016AddingFlock} for further details.
A related approach to influences a flock's direction is discussed by \cite{Ben-Shahar2014DirectionSwarms}, where the authors propose a method to optimally cycle between different leading agents to steer the flock in a desired direction.
The authors provide additional guarantees that no collisions will occur and that all agents will remain connected to the flock.
}

\begin{figure}[ht]
    \centering
    \includegraphics[width=0.7\linewidth]{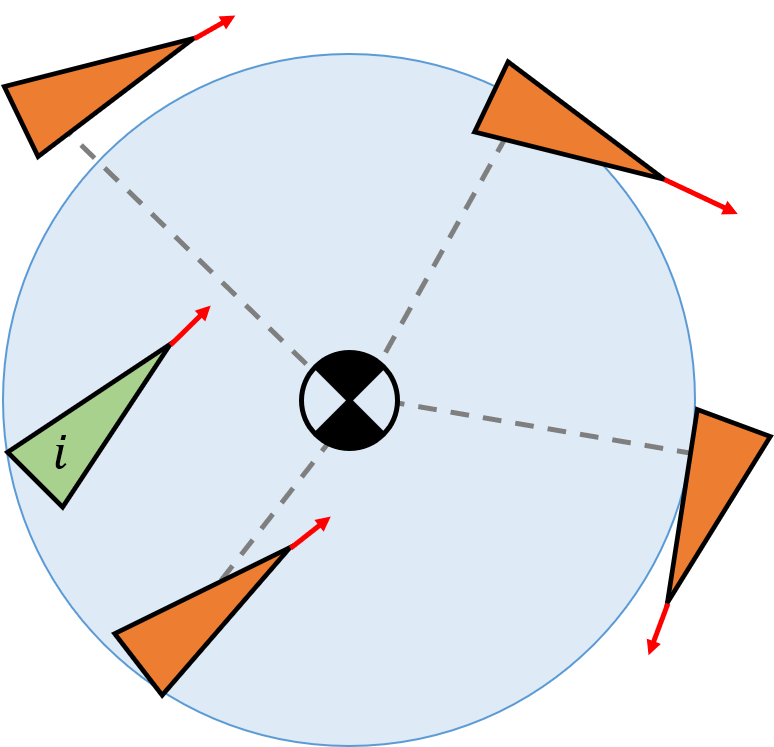}
    \caption{Agent $i$, in green, is constrained to remain within a disk positioned at its neighborhood center.}
    \label{fig:diskFlocking}
\end{figure}

Inspired by Reynolds flocking rules and the constraint-driven paradigm for control, \cite{Beaver2020BeyondFlocking} propose\blue{s} a set of flocking rules over a planned horizon\blue{, which} achieve cluster flocking by: (1) minimizing energy consumption, (2) staying near the neighborhood center, and (3) avoiding collisions. Condition 2 (aggregation) is imposed with the constraint
\begin{equation}
    \Big|\Big|\mathbf{p}_i(t) - \frac{1}{|\mathcal{N}_i(t)| - 1} \sum_{j\in\mathcal{N}_i(t)\setminus\{i\}}\mathbf{p}_j(t)\Big|\Big| \leq D,
\end{equation}
for some distance $D$ much greater than the diameter of any agent, and for $|\mathcal{N}_i(t)| > 1$. This approach is visualized in Fig. \ref{fig:diskFlocking}.

The proposed constraint confine\blue{s} each agent within a diameter $D$ disk positioned at their neighborhood center.
The intuition is that the agents may move freely within the disk; however, their velocity cannot vary dramatically from the average velocity in their neighborhood for long periods of time. These rules yield velocity consensus asymptotically when $\mathcal{N}_i(t)$ is forward-invariant.
In a more recent effort, \cite{Beaver2020Energy-OptimalConstraints} proposes a method for a constraint-driven agent to generate an optimal control policy in real-time.
This is an important next step in real-time optimal control of physical flocks.

\section{Line Flocking} \label{sec:line}

\begin{figure}[ht]
    \centering
    \includegraphics[width=0.7\linewidth]{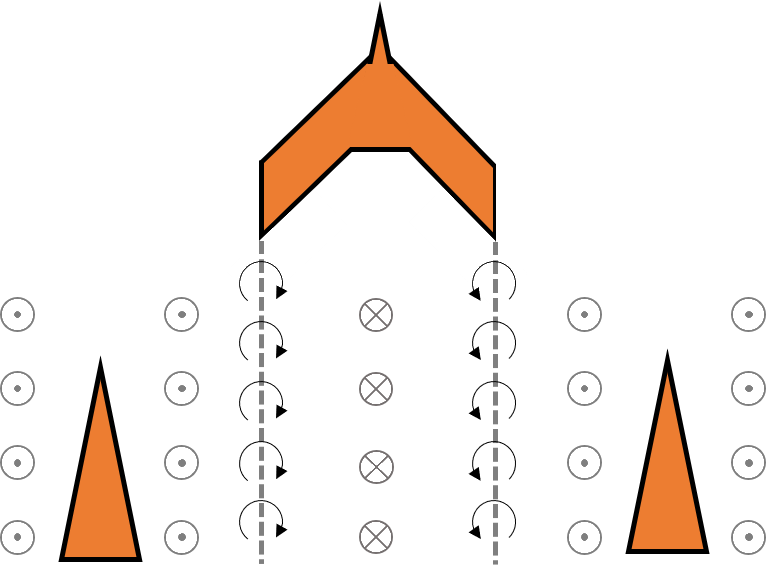}
    \caption{The lead agent induces upwash and downwash in its wake due to its trailing wing vortices, and the following agents exploit the upwash to induce lift and reduce energy consumption.}
    \label{fig:lineFlocking}
\end{figure}

In this section, we review literature related to line flocking, which is a naturally occurring phenomenon commonly found in large birds (such as geese) that travel in a vee, jay, and echelon formations over long distances.
It has long been understood that saving energy is a significant benefit of flying in such formations; see \cite{Cutts1994EnergyGeese,Mirzaeinia2020AnPurposes}.
In aerial systems, the main energy savings comes from upwash, i.e., trailing regions of upward momentum that can be exploited by birds to induce lift and \blue{reduce energy consumption}. 
\blue{Upwash} is illustrated in Fig. \ref{fig:lineFlocking}.
Similar benefits have been found in terrestrial and underwater vehicles, where a leader may create a low-pressure wake \blue{that} reduce\blue{s} the overall drag force imposed on the following vehicles. 

In this context, the most straightforward method to achieve line flocking \blue{may be} to generate an optimal set of formation points based on the drag, wake, and upwash characteristics of each agent. This effectively transforms the line flocking problem into a formation reconfiguration problem, where each agent must assign itself to a unique goal and reach it within some fixed time, as is the case in \cite{Nathan2008V-likeBirds}. However, a formation reconfiguration approach generally requires the formation to be computed offline and does not necessarily consider differences between individual agents (e.g., age, weight, size, and efficiency) or environmental effects. Although formation reconfiguration algorithms have rich supporting literature, they are beyond the scope of this paper. For recent reviews of formation control see \cite{Oh2015,Oh2017}.

Another approach to line flocking is modeling the aerodynamic and hydrodynamic interactions between agents so that they may dynamically position themselves \blue{to save energy}. This \blue{is the approach taken} by \cite{Bedruz2019DynamicOptimization}, \blue{where the authors determine the optimal following distance behind each agent using} computational fluid dynamics simulations 
\blue{of} wheeled mobile robots. 
This \blue{is} extended in \cite{Bedruz2019DesignLogic}, where the authors propose a fuzzy logic controller to maximize the effect of drafting.
The authors validate \blue{these} controllers in simulations and experiments \blue{using} wheeled differential drive robots.

An early approach to capture line flocking behavior in a robotic system with model predictive control is explored in \cite{Yang2016LoveControl}. In this work, the authors \blue{seek} to maximize velocity matching and upwash benefits for each agent $i\in\mathcal{A}$ while minimizing the field of view occluded by leading agents.
\blue{Using simulations, the authors demonstrate} \blue{emergent} line flocking behavior, \blue{and show that} a vee formation consistently emerges independent of the flock's initial conditions.

As a next step toward optimal line flocking, an analysis of the effect of upwash on energy consumption in fixed-wing UAVs is presented by \cite{Mirzaeinia2019EnergyReconfiguration}.
The authors \blue{show} that the front and tail agents in a vee formation have the highest rate of energy consumption in the flock.
This implies that the lead or tail agents become the limiting factor in the total distance traveled by the flock.
The authors propose a load balancing algorithm based on a root-selection protocol, where the highest-energy agents replace the lead and tail agents periodically.
The authors then demonstrate, in simulation, that periodic replacement of the lead and tail agents significantly increases the total distance \blue{travelled by} the flock.

A final facet of line flocking is the effect of environmental disturbances, such as turbulence and currents. Energy-optimal flocking in the presence of strong background flows is investigated by \cite{Song2017Multi-vehicleFlows}.
The authors \blue{calculate} an energy-optimal reference trajectory, $\mathbf{x}_r(t)$, \blue{that the centroid of the flock must track}. 
To generate this trajectory, the authors approximate the flock as a point mass at the centroid and seek to minimize its energy consumption in the presence of a background flow, \blue{denoted} $\mathbf{U}(\mathbf{p}, t)$, where $\mathbf{p}\in\mathbb{R}^2$ is a position.
The normalized rate of power consumption of an agent is given by
\begin{equation}
    P(\mathbf{p}_r(t), t) = \frac{||\mathbf{v}_r(t) - \mathbf{U}(\mathbf{p}_r(t), t) ||^3}{v_{\max}^3},
\end{equation}
and the authors proposed the optimal planning problem
\begin{align*}
    \min_{t^0, t^f, \mathbf{p}_r(t)} & \int_{t^0}^{t^f} P( \mathbf{p}_r (t), t) dt\\
    \text{subject to: } & \mathbf{p}_r(t^0) = \mathbf{p}^0, \mathbf{p}_r(t^f) = \mathbf{p}^f,\\
    & ||\mathbf{v}_r(t)|| \leq v_{\max}, \\
    t_{\min} \leq t^0 < t^f \leq t_{\max}.
\end{align*}
The authors \blue{assume} that the background flow dominates the energy consumption of the agents, and therefore a tight cluster of agents closely approximates the energy-optimal trajectory traced out by the center of the flock.

\section{Pareto Front Selection} \label{sec:pareto}

An essential consideration in multi-objective optimal control is in the trade-off between each of the individual objectives.
This can be observed, for example, in the trade-off between neighborhood centering and velocity alignment in Reynolds flocking.
This trade-off can be explored by finding Pareto-efficient outcomes.
An outcome is Pareto-efficient if no individual term in the cost function can be increased without decreasing the value of any other term; see \cite{Malikopoulos2015}.
The set of all Pareto-efficient outcomes is called the \emph{Pareto frontier}.
After establishing a Pareto frontier, the most desirable outcome can be selected as the Pareto-optimal control policy; see \cite{Malikopoulos2016TAC}.
Pareto frontier generation \blue{is} explicitly discussed in terms of control by \cite{Kesireddy2019GlobalNSGA-III}, who notes that almost all optimal flocking algorithms apply arbitrary weights to the components of multi\blue{-}objective flocking problems.
\blue{The authors present three cooperative evolutionary algorithms, each of which generates a family of control policies that constitutes a Pareto frontier with respect to Reynolds flocking rules.}
\blue{This approach could be used to augment existing evolutionary algorithm approaches, e.g., \cite{Viragh2016Self-organizedEnvironments,Vasarhelyi2018OptimizedEnvironments}, to ensure the resulting control strategies are approximately Pareto optimal.}


\blue{The Pareto frontier is exhaustively searched by} \cite{Hauert2011ReynoldsRate} \blue{to determine the optimal trade-off of} design parameters on flocking performance for a group of UAVs.
The authors note that, due to the hardware limitations, designers must weigh the cost of enhanced communication range versus the maximum turning rate for each agent.
The authors explore this trade-off by exhaustively exploring the design space \blue{to} calculate the resulting heading angle (velocity alignment) and relative drift (flock centering) errors.
Using extensive simulation data, the authors construct the Pareto frontier of optimal design choices.
Finally, to validate their analysis, the authors conduct a set of \blue{four} outdoor experiments using 10 UAVs.

Recent work by \cite{Zheng2020AnTeams} examines the trade-off between flocking performance and privacy. The authors describe a system that follows Reynolds flocking rules guided by a leader robot.
The system is observed by a discriminator \blue{that seeks} to determine which agent is the leader.
The authors propose a genetic algorithm that co-optimizes the flocking controller parameters and the discrimination function.
The \blue{authors present a frontier} of \blue{parameters that influence} flocking performance and leader detectability, \blue{and select the Pareto} optimal \blue{values that yield the best performance for several different leader trajectories}.

\section{Considerations for Physical Swarms} \label{sec:physical}

As the number of agents in a flock increases, the amount of inter-agent communication required may become a significant energy and performance bottleneck.
This has motivated several approaches to minimize the cyberphysical costs incurred by each agent by either reducing the amount of communication required, explicitly including communication cost into an agent's objective function, or \blue{purposely} breaking communication links with a subset of neighboring agents.
In the following subsections, we explore these approaches and discuss their potential value to optimal flocking.

\subsection{Reducing Cyberphysical Costs} \label{sec:cyberPhysical}

The cost of communication \blue{is} explicitly included by \cite{Li2013FlockingOptimization} as part of a holistic cyberphysical approach \blue{to flocking}.
To account for environmental and inter-agent communication disturbances, the authors calculate the probability of communication errors as a function of \blue{the} physical antenna properties.
\blue{The authors then select a maximum failure risk threshold, which determines the maximum separating distance allowed between neighboring agents.}
Based on the collision avoidance constraint and maximum communication distance, each agent $i\in\mathcal{A}$ determines a minimum and maximum distance to every neighbor $j\in\mathcal{N}_i(t)$. 
Agent $i$ selects the optimal separating distance, within these bounds, to minimize a combination of communication error and a crowding penalty.
\blue{In \cite{Li2017FlockingPerspective},} the authors propose an adaptive controller \blue{for} the optimal separating distance, and \blue{they} extend the analysis to include both near and far-field communication.

A control method for preserving agent connectivity while minimizing the number of neighbors \blue{is} \blue{proposed by} \cite{Zavlanos2009}.
In this formulation, agents receive a number of communication hops from their neighbors\blue{, which} they use to estimate the communication graph diameter.
\blue{This information is used by each agent to break the maximum number of communication links while guaranteeing that the communication topology remains connected.}
Graph topology \blue{is} explicitly linked with antenna power in \cite{Dolev2010Bounded-HopSwarms}, where the agents \blue{seek} to minimize communication power while guaranteeing a minimum global graph diameter.
This work \blue{is expanded upon} in \cite{Dolev2013Bounded-hopSwarms}, where agents appl\blue{y} a gossip algorithm to achieve global information about the system trajectory. 
Communication hop approaches have \blue{already} been successfully used in more centralized and structured swarm problems, particularly pattern formation; see \cite{Rubenstein2012, Wang2020ShapeSwapping}.
Although these \blue{methods} are not directly applicable to swarm systems, a similar approach may be beneficial to ensure that all agents satisfy Definition \ref{def:clusterFlocking} while minimizing communication costs between agents.
Finally, \cite{Chen2012TheSynchronization} \blue{seeks to determine} the minimum possible communication distance \blue{that} guarantees \blue{agents under the Vicsek flocking model; see \cite{Vicsek1995NovelParticles},} convergence to velocity consensus. 
The authors show that if the position and orientation \blue{of the agents are} randomly and uniformly distributed in $[0, 1]^2\times[-\pi, \pi]$, \blue{then} the minimum possible communication distance is $\sqrt{\frac{\log{N}}{\pi N}}$.
This provides a lower bound on \blue{the minimum} communication energy cost for the flock \blue{under the distance-based neighborhood metric}.

\cite{Camperi2012SpatiallyModels} \blue{studies} the stability of a flock when noise and external perturbations are introduced.
The authors \blue{seek} to optimize the \blue{resistance to fragmentation} of a large swarm of Vicsek agents in $\mathbb{R}^3$ by changing the neighborhood topology.
The authors note that, as \cite{Ballerini2008InteractionStudy} found, a $k-$nearest or Voronoi neighborhood topology leads to more stable flocking while reducing the \blue{size of each agent's neighborhood}. 
This has significant implications in how the selection of a neighborhood topology affect\blue{s} the energy cost of communication.

\cite{Zhou2017DistributedOptimization} proposes to minimize communication and computational costs \blue{of MPC approaches} by screening out neighbors that do not negatively impact the objective function of agent $i\in\mathcal{A}$.
The authors appl\blue{y} MPC to a discrete-time flocking system with the $\alpha$-lattice objective \eqref{eq:alphaLattice} and a control penalty term \eqref{eq:controlCost}.
In this case, given a desired distance $d>0$, agent $i$ construct\blue{s} the screened neighbor sets
\begin{align}
    \mathcal{S}_i^1(t) = \{j\in\mathcal{N}_i(t)\setminus\{i\} ~:~ ||\mathbf{s}_{ij}(t)|| > d, \notag\\ \mathbf{s}_{ij}(t)\cdot\dot{\mathbf{s}}_{ij}(t) \geq 0 \},\\
    \mathcal{S}_i^2(t) = \{j\in\mathcal{N}_i(t)\setminus\{i\} ~:~ ||\mathbf{s}_{ij}(t)|| < d,\notag\\ \mathbf{s}_{ij}(t)\cdot\dot{\mathbf{s}}_{ij}(t) \leq 0 \},
\end{align}
where $\mathcal{S}_i^1(t)$ consists of neighbors further than $d$ and moving away, and $\mathcal{S}_i^2(t)$ consists of neighbors closer than $d$ and moving closer.
Thus, agent $i$ must only consider $j\in\mathcal{S}_i^1(t)\bigcup\mathcal{S}_i^2(t)$ when \blue{communicating and} planning.


Another approach to reduc\blue{e} communication and computational cost\blue{s} is to perform sparse planning updates \blue{using} event-triggered control.
\cite{Sun2019FlockingControl} propose\blue{s} an update rule for flocking \blue{systems using the potential field approach} with time delays.
A continuously differentiable and bounded function $\tau(t)$ acts as a time delay on all position measurements.
The authors let the portion of the control input that achieves velocity consensus for agent $i\in\mathcal{A}$ be constant over an interval $[t_1, t_{2})$.
Then\blue{,} the authors propose an error function that the agent uses to update the potential field \blue{(i.e, collision avoidance, component)} of its controller.
\blue{In particular, the event requires} global knowledge of the average agent velocity, communication graph Laplacian, and a Lipschitz bound on the agent dynamics.
The authors prove that, under this event-triggered scheme, the agents converge to steady-state flocking behavior and the system \blue{is} free of Zeno, i.e., chattering, behavior.
This is a promising result for reducing the computational burden on agents, and the development of a decentralized triggering function is a promising area of research.

\subsection{Flocking as a Strategy} \label{sec:strategy}
As we begin to deploy robotic swarm systems in situ, it is crucial to consider when \blue{cluster flocking} is an optimal strategy for a swarm. 
\blue{
Cluster flocking as a system-level strategy has recently been explored using a reinforcement learning approach by \cite{Hahn2019EmergentLearning, Hahn2020ForagingLearning}.
In both cases a swarm of selfish agents is trained using the \emph{parameter sharing} learning technique.
Under the parameter sharing approach, a single neural network is trained from which all agents derive their actions.
Furthermore, during each learning episode a single agent is selected to update the shared policy using only locally available information.
After each learning episode, the shared policy is updated across all agents and a new learning agent is selected.
}

\blue{
\cite{Hahn2019EmergentLearning} proposes an approach similar to \cite{Morihiro2006EmergenceLearning}, which yields an emergent swarming behavior for predator avoidance.
The authors fix the control policy of the predator such that it always moves toward the nearest agent.
In the case that multiple agents are equally close, the predator selects an agent at random.
This switching behavior presents an opportunity for the agents to confuse the predator by forcing it to frequently change which agent it is moving toward.
The authors also include a minimum dwell time before the predator is able to switch the agent that it is following.
The learning agent is rewarded for every time step that it avoids the predator, and is heavily penalized if it comes within a fixed distance of the predator.
Under this setup, and without including any explicit flocking rules, a cluster flocking behavior emerges where the agents dynamically form and break apart clusters in an attempt to change the agent that the predator is following.
}

\blue{
Parameter sharing is also employed by \cite{Hahn2020ForagingLearning} to solve the centroid tracking problem using a swarm of fixed-speed agents.
The system-level objective is to track a moving food source, while each agent is rewarded proportionally to its distance from the target.
In addition, the speed of an agent is reduced by half whenever it collides with a neighboring agent.
Finally, the authors compare the performance of their learned algorithm to Reynolds flocking, a heuristic strategy, and a control policy of taking random actions.
The authors show through simulation, and without incorporating any flocking rules, their learning agents outperform the other methods when tracking the reference trajectory.
Additionally, a global alignment behavior emerges, i.e., the agents aligned themselves with their neighbors to avoid collisions and approach the target when it is outside of their maximum sensing distance.
}

Line flocking as a strategy has been explored as a trade-off between the energy savings of flocking and the energy cost of rerouting to join a flock.
Significant research effort has gone toward the rendezvous problem, that is, given a set of agents with distinct origins and destinations, when is it optimal for agents to expend energy in order to form an energy-saving flock.
This has primarily been explored through the lens of air traffic management, where commercial aircraft may rendezvous to form flocks between origin and destination airports given a takeoff and landing window.
A centralized approach to the rendezvous problem \blue{is} presented in  \cite{Ribichini2003EfficientSystems}, where the authors prove several properties of energy-optimal rendezvous for two agents.
The two-agent case \blue{is} further explored in \cite{Rao2006OptimalBeneficial} for minimum-time graph traversal.
The effect of wind and environmental factors \blue{is determined} in \cite{Marks2018IdentificationWind}, where the authors use historical traffic and environmental data to show a 5-7\% increase in fuel economy resulting from coordination.
A flocking protocol for selfish agents \blue{is} presented in \cite{Azoulay2019UAVEnvironments}, and the air traffic routing problem \blue{is} extensively explored in \cite{Kent2015OptimalFlight} and \cite{Verhagen2015FormationRouting}.
\blue{Line f}locking as a strategy \blue{is} also explored in the context of passenger vehicle eco-routing by \cite{Fredette2017Fuel-SavingControl}.
In this approach, the author adapt\blue{s} Reynolds flocking rules to a two-lane highway with the objective of minimizing vehicle fuel consumption while maintaining a desired velocity subject to the physical parameters \blue{that} describ\blue{e} each vehicle.
This result\blue{s} in each vehicle approaching its desired speed while dynamically forming and exiting flocks under a centralized control scheme.

\section{Outlook and Research Directions} \label{sec:outlook}

In the past twenty years, a rich literature on the control of flocking systems has been produced.
Control algorithms that implement variants of Reynolds rules have proven rigorous guarantees on their steady-state behavior.
Recently, control algorithms that optimally implement these rules have been demonstrated in simulation and large-scale outdoor flight tests.
\blue{Therefore, optimal Reynolds f}locking will seemingly be driven by advances in decentralized control, robust control, and long-duration autonomy in the future.
However, some application areas, such as mobile sensor networks, have criticized Reynolds flocking as a novelty that does not necessarily have advantages in terms of performance or ease of implementation; see \cite{Albert2018Survey:Tracking} for a motivating example.

\blue{For this reason we have proposed a new paradigm to understand} the nature of flocking. 
As we demonstrated, there is a distinction in the natural world between cluster and line flocking.
We wish to strengthen this distinction, and to that end, we propose a \blue{flocking taxonomy that partitions} the literature into line and cluster flocking \blue{as we present in Fig.  \ref{fig:flockingCategories}}.
We have also presented several types of cluster flocking, defined by the\blue{ir} system-level objective, \blue{which} have \blue{all} been conflated using the nebulous term ``flocking'' \blue{in the past}.
\blue{In fact, } due to the nature of engineering systems, new types of cluster flocking have already emerged that have no natural counterpart.
For this reason, we believe that precisely classifying \blue{cluster flocking problems based on the system-level objective} will be essential to advancing the research frontier of flocking as a desirable emergent behavior.
\blue{Furthermore, classifying the line flocking literature around the exploited energy-saving mechanism is a practical step to enhance collaboration in these domains and to continue driving line flocking research forward. }

\blue{We believe that there will be at least four significant challenges in the near future on systems with optimal flocking.}

\blue{First, the relationship between information and performance in decentralized systems will be an area that would require to overcome significant challenges.
As robotic flocks emerge they will generate information and make decisions at a rate that is infeasible for a central controller or database to manage.
Thus, the ability for each agent to identify what local information is of high value will become increasingly important, particularly for applications in isolated or spatially distributed settings.
}

\blue{Second, while great strides have been made in terms of generating optimal trajectories, there is a clear gap for holistic approaches that includes the communication, sensing, and computational costs necessary for fully autonomous flocking.}

\blue{Third, there is no rigorous mathematical approach to predict the emergence and stability of flocking behavior in decentralized system.
For this challenge we propose} 
constraint-driven optimal control \blue{as a} natural \blue{choice} to \blue{provide guarantees on emergent flocking behavior}.
Under this design paradigm, it is possible to achieve rigorous guarantees on the safety and tasks imposed on agents as they travel along energy-minimizing \blue{trajectories}.
There has already been some initial exploration into Reynolds flocking, e.g.,see \cite{Ibuki2020Optimization-BasedBodies}, and systems with limited communication range under disk flocking; see \cite{Beaver2020BeyondFlocking}.
These approaches have also shown a capacity for generating emergence in relatively simple multi-agent systems, e.g., see \cite{Notomista2019Constraint-DrivenSystems}, and the imposed constraints provide guarantees on agent behavior to neighbors and the system designer.
Moving forward, we expect that by applying similar solution methods to those used in the past, e.g., see \cite{Jadbabaie2003,Tanner2007}, we may provide guarantees on the behavior of many types of cluster flocking agents.

Finally, heterogeneity in cluster and line flocking will be essential as optimal flocking algorithms \blue{are realized on} physical systems, where it is \blue{effectively} impossible for any two robots to have identical physical properties and performance capabilities.
Heterogeneity of agent properties is particularly important in the \blue{biological} line flocking literature, where the variable size, wingspan, metabolism, and age of flock members significantly affects the system's overall energy savings; see \cite{Mirzaeinia2020AnPurposes}.
\cite{Prorok2017TheSwarms} also show\blue{s} that for general swarm systems, an increase in agent diversity expand\blue{s} the feasible solution space for each agent's control action.
This may be beneficial \blue{for designing} robust \blue{systems}, especially for applications related to emerging transportation systems; however, it may also increase the difficulty of finding an optimal solution.
Future flocking research ought to consider diversity in agent properties and behaviors to exploit the full benefits of swarm intelligence.

\section*{Acknowledgement}
The authors would like to thank \blue{Dr}. Bert Tanner for \blue{his} insightful remarks and suggestions.

\bibliography{idsPubs,mendeley,other}

\end{document}